\documentclass[conference]{IEEEtran}
\usepackage{stmaryrd}
\usepackage{amsfonts}

\usepackage{graphicx,times,amsmath} 
\usepackage{algorithm}
\usepackage{algorithmic}

\usepackage{threeparttable}
\usepackage{dcolumn}
\usepackage{multirow}
\usepackage{multicol}
\usepackage{booktabs}
\usepackage{bm}
\usepackage{lineno}

\usepackage{url}
\usepackage{subfigure}
\usepackage{color}

\IEEEoverridecommandlockouts    

\begin{document}

\title{\ \\ \LARGE\bf A Framework of Transferring Structures Across Large-scale Information Networks ~\thanks{Shan Xue is with the Lab of Decision Systems \& e-Service Intelligence (DeSI), Centre for Artificial Intelligence (CAI), School of Software, Faculty of Engineering and Information Technology (FEIT), University of Technology Sydney, Australia, and is affiliated with the School of Management, Shanghai University, China (Shan.Xue@student.uts.edu.au).} \thanks{Jie Lu and Guangquan Zhang are with the Lab of Decision Systems \& e-Service Intelligence (DeSI), Centre for Artificial Intelligence (CAI), School of Software, Faculty of Engineering and Information Technology (FEIT), University of Technology Sydney, Australia (\{jie.lu, guangquan.zhang\}@uts.edu.au).} \thanks{Li Xiong is with the School of Management, Shanghai University, China (xiongli8@shu.edu.cn).}  }

\author{Shan Xue, Jie Lu, Guangquan Zhang, and Li Xiong}


\maketitle

\begin{abstract}

The existing domain-specific methods for mining information networks in machine learning aims to represent the nodes of an information network into a vector format. However, the real-world large-scale information network cannot make well network representations by one network. When the information of the network structure transferred from one network to another network, the performance of network representation might decrease sharply. To achieve these ends, we propose a novel framework to transfer useful information across relational large-scale information networks (FTLSIN). The framework consists of a 2-layer random walks to measure the relations between two networks and predict links across them. Experiments on real-world datasets demonstrate the effectiveness of the proposed model.

\end{abstract}


\section{Introduction}\label{sec1}

\PARstart{I}{nformation} networks \cite{newman2010networks} is a kind of structure-based data, \textit{e.g.,} academic citation networks, which employ network topologies to save a part of information directly. Large-scale information network ranges the size from hundreds of nodes to millions and billions of nodes \cite{tang2015line}. The large volume of nodes make complex connections over the network and contains complex data structures than normal information networks. To fully analyze such kind of information networks is a quite challenging problem especially in machine learning.

 Network representation, also known as network embedding, allows analyzing the network structure and mining the information behind the structure in a machine learning perspective \cite{wang2017community,huang2017label}. By generating a latent representation space in relatively low dimensions from the interactions in high dimensions, network representation inputs a structured data of graph \cite{Pan:2015:FBM:2824238.2824407} and outputs the embeddings of the graph in a specific dimensional space. It guarantees the correspondence between community structure in the input graph and its embeddings. The main advantage of network representation is that the learned representations encode community structure, so it can be easily exploited by simple and standard classifiers \cite{bhagat2011node}.

 For the purpose of using network analysis on structured data, a series of models are proposed based on DeepWalk \cite{perozzi2014deepwalk}, which trains a natural language model on the random walks generated by the network structure. Denote a random walk $w_{v_s}$ that starts from a root node $v_s$, DeepWalk slides a window in a length of $2w+1$, and maps the central node $v_i$ to its representation $f(v_i)$. Hierarchical Softmax factors out the probability distributions $Pr(v_{i\pm d}|f(v_i))$, where $d=\{1,\cdots,w\}$, corresponding to the paths staring at $v_i$ and going over all other nodes in the random walk. The representation $f$ is updated to maximize the probability of $v_i$ co-occurring with its context $\{v_{i\pm d},~d=\{1,\cdots,w\}\}$. Random walk based DeepWalk shows promising results on large-scale network representation if the datasets have a satisfying structure.

 LINE and Node2Vec are the other two structure-based network representation models that improves the performance of DeepWalk. LINE \cite{tang2015line} preserves both local and global network structure by first-order proximity and second-order proximity respectively and suitable for all kinds of networks, \emph{i.e.,} directed and undirected networks and weighted and unweighted networks. Node2Vec \cite{grover2016node2vec} explores the diverse neighborhoods of nodes in a biased random walk procedure with search bias $\alpha$.

 Above mentioned models are inspired from recent advancements in unsupervised feature learning and the language modeling from sequences of words to vectors or networks. They contributed to the network analysis by modeling a stream of short random walks. Different from traditional representation learning, the latent feature learning of network representation captures neighborhood similarity and community membership in topologies. The potential of these models in real-world scenarios is their good performance in large heterogeneous networks.

\begin{figure*}[!t]
\centering
\includegraphics[width=0.8\linewidth]{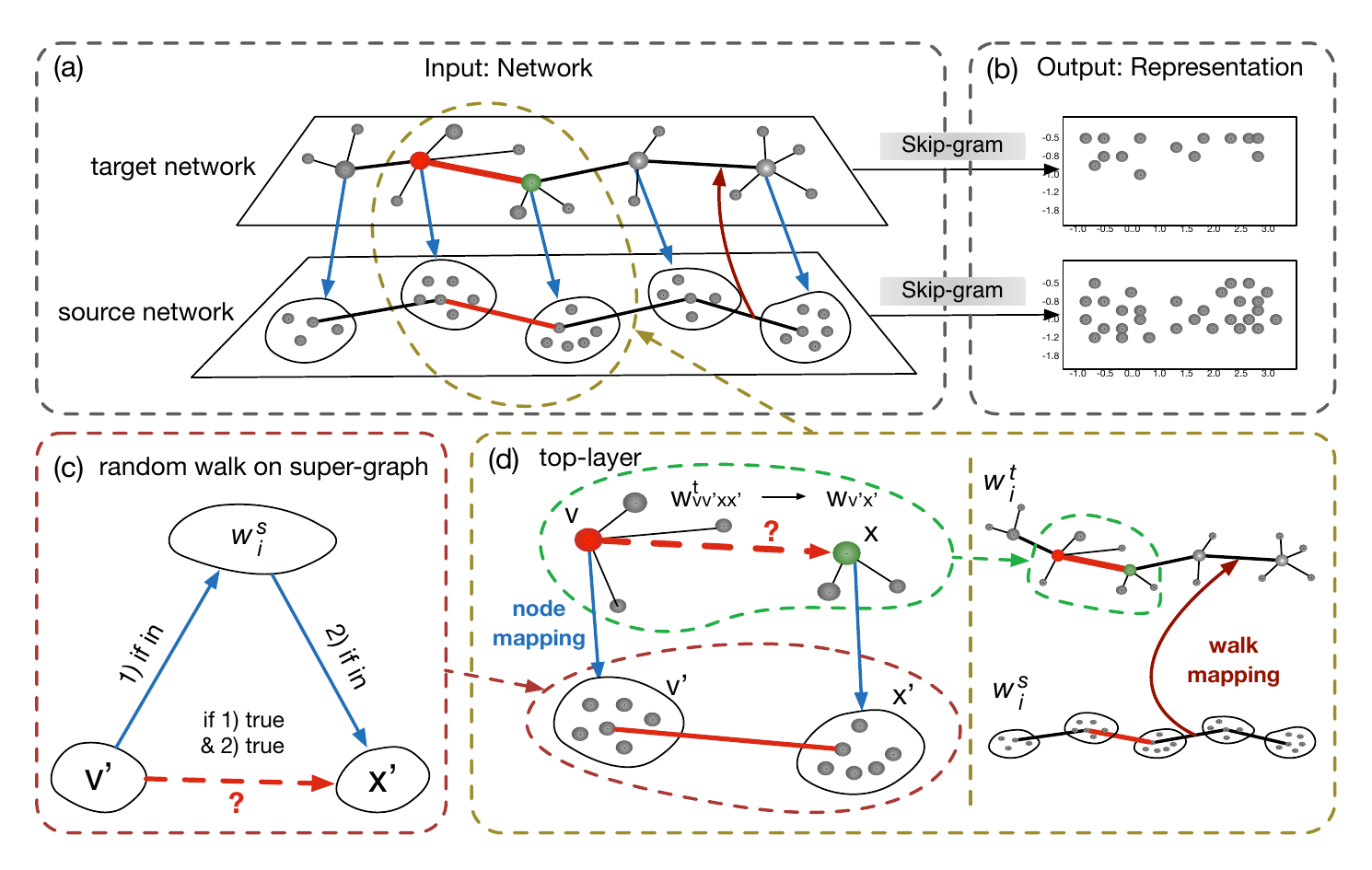}
\caption{Large-scale Information Network Structures Transfer Framework}
\label{fig:CDNRFramework}
\end{figure*}

 When one network lacks of connections, the existing network representation methods cannot perform well in domain-specific network representations. Meanwhile, the performance of the domain-specific network representations methods decreases sharply when transferred from one network to another relational network. If we develop a framework that successfully combines the advantages of network representation and domain adaptation, structure transfer will extremely benefit to real-world large-scale information network representations.

 In this paper, we consider the following challenges of developing the framework for transferring the network structures across large-scale information networks.

\begin{itemize}
  \item \textbf{Challenge 1}: How to effectively predict links between nodes across relational networks for the purpose of improving performance of network representation in the target network?
  \item \textbf{Challenge 2}: How to transfer the random walks in the source network to the target network based on the similarity measurement achieved in Challenge 1?
\end{itemize}

To this end, we propose a framework of transferring structures across large-scale information networks (FTLSIN) which implements an unsupervised feature learning for scalable networks. Our framework is built by a 2-layer random walks to generate a neighborhood of nodes in the target network with a secondary from the learned walks in the source network which measures the similarity and predicts links across networks. Experimental results on real-world datasets empirically demonstrate that our framework achieves better performances compared to the state-of-the-art network representations.

\section{Problem Statement}\label{sec2}

The problem of transferring structures across large-scale information networks is formulated as follows. Suppose we have a source domain $\mathcal{D}_s$ and a target domain $\mathcal{D}_t$, where the source domain has a source networks $G^s=(V^s,E^s)$ with its corresponding label space $Y^s$, and the target domain has a target network $G^t=(V^t,E^t)$ with its label space $Y^t$. Both networks are unweighted. For a cross-domain classification problem $<\mathcal{T}_t,\mathcal{T}_s,(\textbf{x}^t_{test},y^t_{test})>$, we firstly implement a latent feature learning procedure from topology structures of $G^s$ and $G^t$ as a maximum likelihood optimization problem and then learn the labels $Y^t$ in the target domain with standard classifiers as an evaluations of the cross-domain network representations.

In this paper, let $f: V\rightarrow \mathbb{R}^d$ be the mapping function from nodes to feature representation, where $d$ refers to the lower-dimensions of our representation, $f$s are specially designed for the source network and the target network respectively, i.e., $f^s: V^s \rightarrow \mathbb{R}^d$ and $f^t: V^t \rightarrow \mathbb{R}^d$. As a sampling strategy, we define a neighborhood of nodes $N_S(u)\subset V$ for every node in the source network and in the target network, where $N_S(u^s)\subset V^s$, $u^s \in V^s$, $N_S(u^t)\subset V^t$ and $u^t \in V^t$. By predicting the latent feature space $\mathbb{R}^d $, our proposed framework can be applied to any (un)directed and (un)weighted network across domains.

\section{Large-scale Information Network Structures Transfer Framework}\label{sec3}

\subsection{Skip-gram in FTLSIN}

FTLSIN shown in Fig. \ref{fig:CDNRFramework} learns random walks by Skip-gram and outputs the network representations from the input of networks in source domain and target domain, respectively. Skip-gram \cite{mikolov2013efficient} is a language model exploiting word orders in a sequence and assuming that words closer are statistically more dependent or related. We employ the Skip-gram architecture to FTLSIN which treats the nodes in a sequence and fully uses the structures to make network analysis.

Given a current node $u^s$ in the source network within a certain window, we have a FTLSIN Skip-gram for source networks by maximizing the following log-likelihood function of $f^s$ in observing a neighborhood of $N_S(u^s)$:

\begin{equation}\label{eq:source_skip-gram}
    \max_{f^s}~~\sum_{u^s\in V^s}\log{Pr(N_S(u^s)|f^s(u^s))}
\end{equation}

Given a node $u^t$ in the target network with a certain window, we have a FTLSIN Skip-gram for the target network by maximizing the following log-likelihood function of $f^t$ in observing a neighborhood of $N_S(u^t)$:

\begin{equation}\label{eq:target_skip-gram}
    \max_{f^t}~~\sum_{u^t\in V^t}\log{Pr(N_S(u^t)|f^t(u^t))}
\end{equation}

Following the standard assumptions of Node2Vec model \cite{grover2016node2vec}, conditional independence and symmetry in feature space are defined in Eqs. (\ref{eq:conditional_independence}) and (\ref{eq:symmetry}), respectively.

\begin{equation}\label{eq:conditional_independence}
    Pr(N_S(u)|f(u))=\prod_{n_i\in N_S(u)}Pr(n_i|f(u))
\end{equation}

\begin{equation}\label{eq:symmetry}
    Pr(n_i|f(u))=\frac{\exp{(f(n_i)\cdot f(u))}}{\sum_{v\in V}\exp{(f(v)\cdot f(u))}}
\end{equation}

In our proposed FTLSIN Skip-gram (see in Algorithm \ref{alg:skip-gram_target}), the network neighborhood strategy applied on the target network can be different from the ones on source networks. Meanwhile, the window length and optimization function $f^s$ and $f^t$ set in FTLSIN Skip-gram also can differ from networks.

\subsection{2-layer Random Walks in FTLSIN}

The FTLSIN 2-layer random walks consists of a bottom-layer random walk and a top-layer random walk. In Fig. \ref{fig:CDNRFramework}, the 2-layer random walks measures the likelihood between super-nodes $\{v',x'\}$ based on its learning of random walk $w^s_i$. The top-layer maps two nodes $\{v,x\}$ in the target network to the corresponding super-nodes $\{v',x'\}$ in a source network within a node mapping procedure and a walk mapping procedure. The Algorithm of the 2-layer random walks in FTLSIN is as shown in Algorithm \ref{alg:CD2LRW}.

Given a random walk of node in either a source network $u^s$ or a target network $u^t$, $u$ is in a fixed length of $l$, i.e., the length of $u^s$ is $l^s$ and the length of $u^t$ is $l^t$. 2-layer random walks allows $l^s$ different from $l^t$.

Let $v_i$ denote the $i$th node in the walk, where the start node is $v_0=u$ and all the nodes in the walk follows the distribution:

\begin{align}\label{eq:random_walk_s}
    P(v^s_i=x^s&|v^s_{i-1}=v^s)\notag\\
    =&\begin{cases}
    	\frac{\pi_{v^sx^s}}{Z}~~&\text{if}~(v^s,x^s)\in E^s\\
    	0~~&\text{otherwise}
    \end{cases}
\end{align}

\begin{align}\label{eq:random_walk_t}
    P(v^t_i=x^t&|v^t_{i-1}=v^t)\notag\\
    =&\begin{cases}
    	\frac{\pi_{v^tx^t}}{Z}~~&\text{if}~(v^t,x^t)\in E^t\\
    	0~~&\text{otherwise}
    \end{cases}
\end{align}
where $\pi_{v^sx^s}$ and $\pi_{v^tx^t}$ are the unnormalized transition probability between nodes $v^s$ and $x^s$, and between nodes $v^t$ and $x^t$; and $Z$ is the normalizing constant.

 \subsubsection{Bottom-layer Random Walk}

 The design of the bottom-layer random walk is for the network representation both in the target network and in the source network. We employ parameters $p$ and $q$ to guide the walk by considering the network neighborhood. In order to determine which node in the neighborhood have a higher probability to be connected into the random walk, the search bias $\alpha$ is employed into Eqs. (\ref{eq:bottom_layer_random_walk_s}) and (\ref{eq:bottom_layer_random_walk_t}):

\begin{equation}\label{eq:bottom_layer_random_walk_s}
    \pi_{v^sx^s}=\alpha_{pq}(t^s,x^s)\cdot w_{v^sx^s}
\end{equation}
where $w_{v^sx^s}$ is the weight on edge $(v^s,x^s)$.

\begin{algorithm}[t]
  \caption{FTLSIN 2-Layer Random Walks}\label{alg:CD2LRW}
  \renewcommand{\algorithmicrequire}{\textbf{Input:}}
  \renewcommand\algorithmicensure {\textbf{Output:} }
  \begin{algorithmic}[1]
  \REQUIRE ~~\\
  $G^t=(V^t,E^t)$: a target network;\\ $G^s=(V^s,E^s)$: a source network.\\
  \ENSURE ~~\\
  $W^t$: a walk set of target network.\\
  	\STATE $W^s \leftarrow$ Apply bottom-layer random walk to process the source network, Eqs. (\ref{eq:random_walk_s})-(\ref{eq:search_bias_s}).
  	\STATE $G^s=(\mathcal{V}^s,\mathcal{E}^s,\mathcal{G}^s,\mathcal{F}^s)\leftarrow$ Samples the source network to a super-graph with super-nodes.
  	\FOR{$w^s_i$ in $W^s$}
	  	\STATE $f_{node}\leftarrow$ Node mapping on $w^s_i$ by Eq. (\ref{eq:fnode}).
	  	\STATE $w_{v'x'}\leftarrow$ Wwalk mapping on $w^s_i$ and $f_{node}$ by Eqs. (\ref{eq:WMWall})-(\ref{eq:WMW1}), where $v',x'\in \mathcal{V}^s$.
  	\ENDFOR
  	\STATE $W^t\leftarrow$ Apply bottom-layer random walk to process the target network, Eqs. (\ref{eq:random_walk_s})-(\ref{eq:search_bias_t}), where $w_{v^tx^t}=w_{v'x'}$.
  	\RETURN $W^t$
  \end{algorithmic}
\end{algorithm}

\begin{align}\label{eq:search_bias_s}
    \alpha_{pq}(t^s,x^s)=
    \begin{cases}
    	\frac{1}{p} ~~&\text{if}~d_{t^sx^s}=0\\
    	1~~&\text{if}~d_{t^sx^s}=1\\
    	\frac{1}{q} ~~&\text{if}~d_{t^sx^s}=2
    \end{cases}
\end{align}
where $d_{t^sx^s}$ is the shortest path between nodes $t^s$ and $x^s$ through node $v^s$.

\begin{equation}\label{eq:bottom_layer_random_walk_t}
    \pi_{v^tx^t}=\alpha_{pq}(t^t,x^t)\cdot w_{v^tx^t}
\end{equation}
where $w_{v^tx^t}$ is the weight on edge $(v^t,x^t)$.

\begin{align}\label{eq:search_bias_t}
    \alpha_{pq}(t^t,x^t)=
    \begin{cases}
    	\frac{1}{p} ~~&\text{if}~d_{t^tx^t}=0\\
    	1~~&\text{if}~d_{t^tx^t}=1\\
    	\frac{1}{q} ~~&\text{if}~d_{t^tx^t}=2
    \end{cases}
\end{align}
where $d_{t^tx^t}$ is the shortest path between nodes $t^t$ and $x^t$ through node $v^t$.

\subsubsection{Top-layer Random Walk}

For the random walk in the top-layer, we define a node mapping procedure and a walk mapping procedure. The node mapping procedure starts from one node $v\in V^t$ in the target network to a node set $v'\in V^s_i$ in the source network. The walk mapping procedure learns from a walk $w^s \in \{W^s_i\}$ in the source network to a new walk $w^t \in {W^t}$ in the target network.

Following the assumption of transfer learning \cite{lu2015transfer}, the scale of networks in the source domain is much larger than the scale of the network in the target domain, \textit{i.e.,} $|V^s_i|\gg|V^t|$ or $|E^s_i|\gg|E^t|$. The node mapping procedure links a node in target network and a set of nodes in source network. Thus, we employs the definition of super-graph and super-node to process the node mapping procedure. Specifically, the node set $v'$ is denoted as a super-node.

A super-graph \cite{guo2014super} is represented as $G=(\mathcal{V},\mathcal{E},\mathcal{G},\mathcal{F})$, where $\mathcal{V}$ is a finite set of graph-structure nodes. $\mathcal{E}\subset \mathcal{V}\times \mathcal{V}$ denotes a finite set of edges, and $\mathcal{F}: \mathcal{E}\rightarrow \mathcal{G}$ is an injective function from $\mathcal{E}$ to $\mathcal{G}$, where $\mathcal{G}$ is the set of single-attribute graphs. A node in the super-graph, represented by a single-attribute graph, is called a super-node.

As above, our node mapping procedure measures the likelihood of a node in the target network and a super-node in the source network, \textit{i.e.,} $f_{node}: v^t \rightarrow V^s_i$.

\begin{align}\label{eq:fnode}
    f_{node}=
    \begin{cases}
    	1 ~~&\text{if}~\deg{(v^t)}=\deg{(V^s_i)}\\
    	0~~&\text{otherwise}
    \end{cases}
\end{align}

In the walk mapping procedure, walk set of target network $W^t$ is jointly determined by the node mapping function $f_{node}$ and weighted random walk kernel \cite{guo2014super} on super-graph in source network. The walks over the source network $W^s=\{w^s_i\}$ is naturally within a super-graph. The edges forming a $w^s_i$ links two super-nodes, as shown in Fig. \ref{fig:CDNRFramework} (d).

\begin{algorithm}[t]
  \caption{FTLSIN Skip-gram}\label{alg:skip-gram_target}
  \renewcommand{\algorithmicrequire}{\textbf{Input:}}
  \renewcommand\algorithmicensure {\textbf{Output:} }
  \begin{algorithmic}[1]
  \REQUIRE ~~\\
  $W^t$: a target random walk set learning from Algorithm \ref{alg:CD2LRW}.\\
  \ENSURE ~~\\
    $f^t:~V^t\rightarrow \mathbb{R}^d$: an optimized mapping function for FTLSIN.\\
  \STATE $f^{t(0)}\leftarrow$ Initialize the target network mapping function.
  	\FOR{$w^t_i$ in $W^t$}
	  	\STATE $f^t\leftarrow$ Apply Eq. (\ref{eq:target_skip-gram})-(\ref{eq:symmetry}) to optimize $f^t$.
  	\ENDFOR
  	\RETURN $f^t$
  \end{algorithmic}
\end{algorithm}

Within the top-layer random walk, an edge weight in a target walk ($w_{v'x'}$ in $w^t_i$)  is formed by two terms in Eq. (\ref{eq:WMWall}). The former term is contributed by the virtual weight in the target network $w_{v'^tx'^t}$, and the latter term is contributed by the learning weight from a walk mapping $w^t_{vv'xx'}$.

\begin{equation}\label{eq:WMWall}
    w_{v'x'}=\beta \cdot w^t_{vv'xx'}+(1-\beta)\cdot w_{v'^tx'^t}
\end{equation}
where $\beta=|V^t|/(|V^s_i|+|V^t|)$.

\begin{equation}\label{eq:WMW1}
\small
    w^t_{vv'xx'}=\max_{f_{walk}} \sum_{W^s_{w_i}\in \mathcal{W}^s}\sum_{\substack{v^s\in v',\\x^s\in x'}}\log P(x^s|v^s)
\end{equation}
where $P(x^s|v^s)=1/d_{v^sx^s}$.

\begin{figure}[!b]
\centering
\subfigure[Source Network: dblp]{\includegraphics[width=0.45\linewidth]{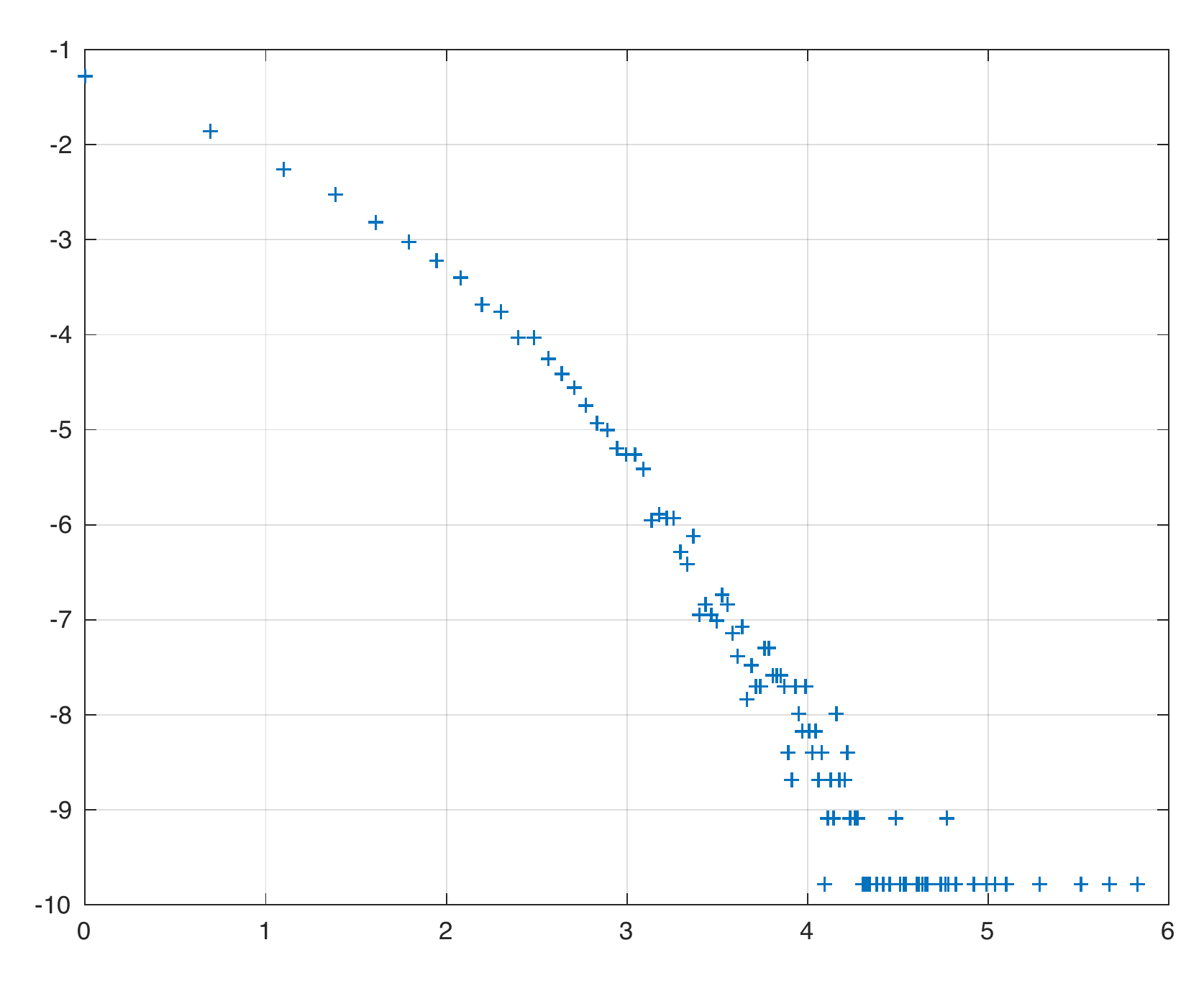}}
\subfigure[Target Network: M10]{\includegraphics[width=0.45\linewidth]{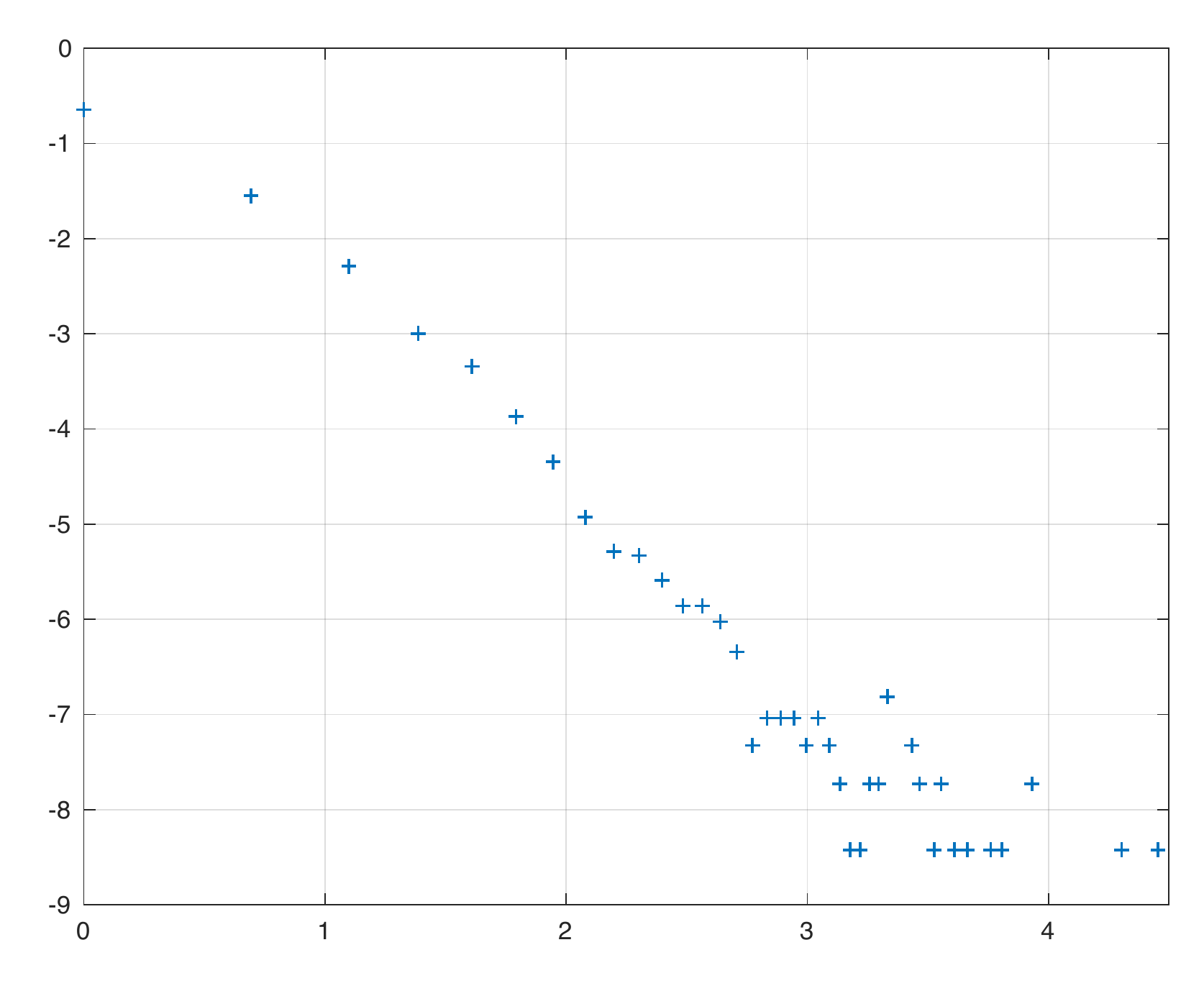}}
\caption{Power-law distributions of the source network and the target network.}
\label{fig:powerlaw}
\end{figure}

\begin{figure}[!b]
\centering
\subfigure[Source Network: dblp]{\includegraphics[width=0.45\linewidth]{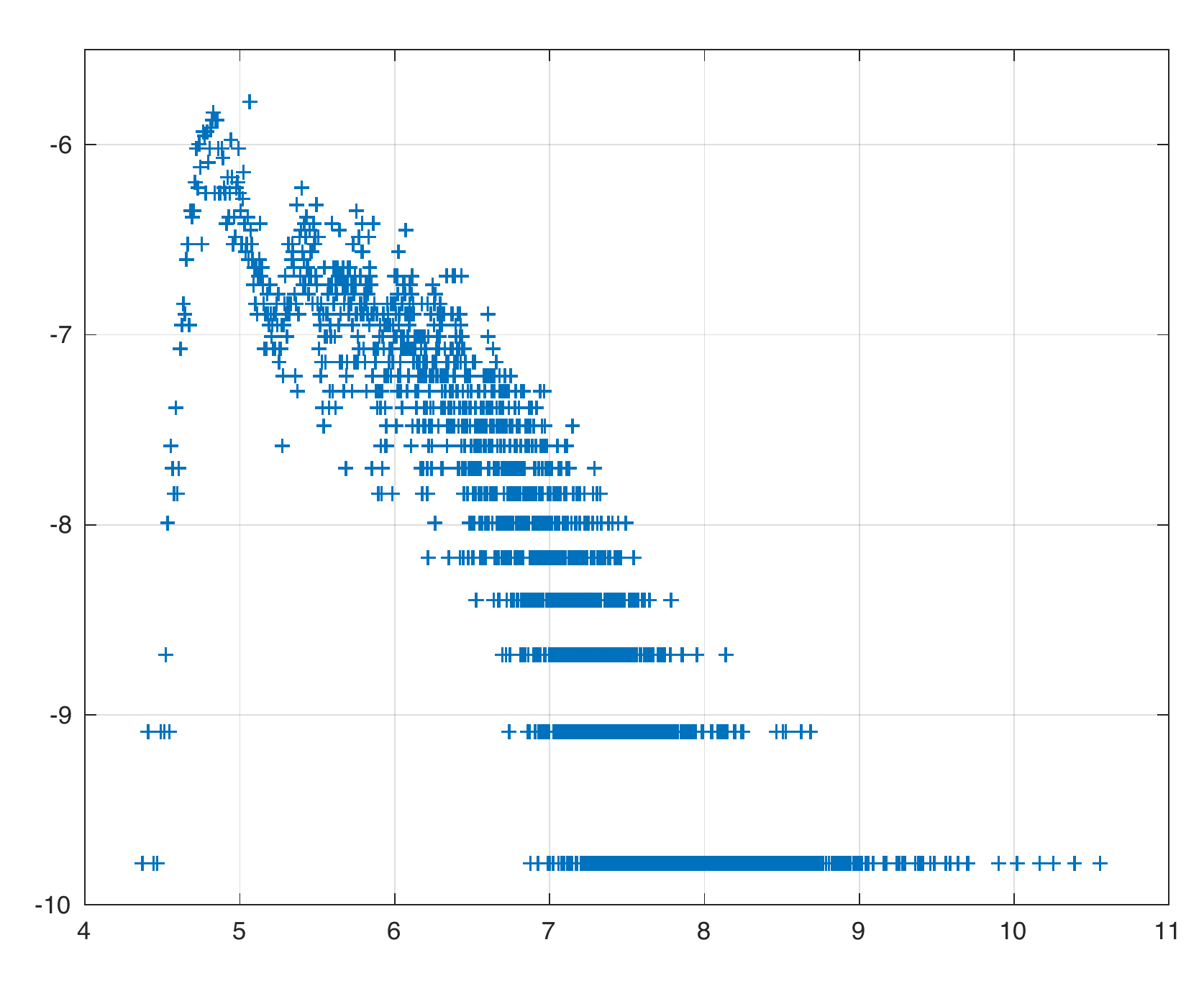}}
\subfigure[Target Network: M10]{\includegraphics[width=0.45\linewidth]{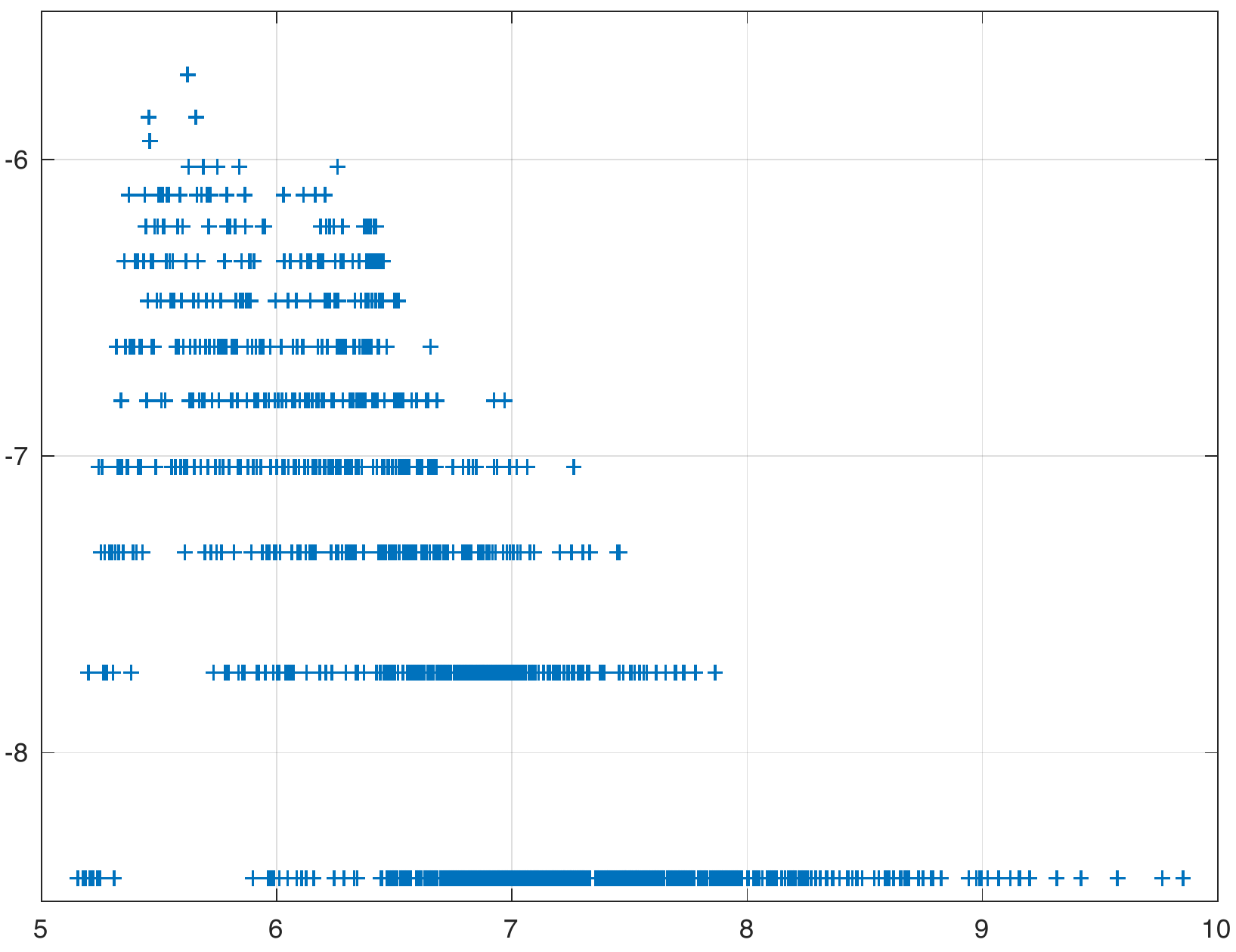}}
\caption{Power-law distributions of the random walks.}
\label{fig:powerlawwalks}
\end{figure}

\section{Experiments}\label{sec4}

\subsection{Datasets}

We select two academic citation networks as the datasets. Both of them are for the multi-class classification problem. Nodes are denoted as papers in these networks.

\begin{table}[!h]
\centering
  \caption{Dataset statistics}
  \label{Table:Statistics of the datasets}
  \begin{tabular}{ccccc}
    \toprule
    \multirow{2}*{Domain} & \multirow{2}*{Network} & Num. of & Num. of & Num. of\\
    & & Nodes & Edges & Labels\\
    \midrule
   Source & DBLP & 60,744 & 52,890 & 4\\
   Target & M10 & 10,310 & 77,218 & 10\\
  \bottomrule
\end{tabular}
\end{table}

\begin{table*}[!t]
\small
\addtolength{\tabcolsep}{+1pt}
\centering \caption{Multi-class classification results on M10 network in target domain}\label{tab:result_target}\vspace{0.1cm}
\begin{tabular}{c||c|c|c|c|c|c|c|c|c|c|c} \hline
 & Model & Statistic & 10\% & 20\% & 30\% & 40\% & 50\% & 60\% & 70\% & 80\% & 90\% \\ \hline \hline
\multirow{8}*{Micro-F1} & \multirow{2}*{DeepWalk} & mean & $0.1758$ & $0.1833$ & $0.1897$ & $0.2049$ & $0.2051$ & $0.2216$ & $0.2236$ & $0.2420$ & $0.2431$ \\
& & variance & $0.0086$ & $0.0100$ & $0.0122$ & $0.0126$ & $0.0128$ & $0.0111$ & $0.0170$ & $0.0133$ & $0.0220$ \\ \cline{2-12}
& \multirow{2}*{LINE} & mean & $0.2338$ & $0.2362$ & $0.2623$ & $0.2821$ & $0.3269$ & $0.3244$ & $0.3561$ & $0.3508$ & $0.4128$  \\
& & variance & $0.0102$ & $0.0170$ & $0.0110$ & $0.0141$ & $0.0150$ & $0.0087$ & $0.0193$ & $0.0184$ & $0.0486$ \\ \cline{2-12}
& \multirow{2}*{Node2Vec} & mean & $0.3342$ & $0.4166$ & $0.4714$ & $0.5213$ & $0.5550$ & $0.5843$ & $0.6216$ & $0.6353$ & $0.6535$  \\
& & variance & $0.0099$ & $0.0110$ & $0.0153$ & $0.0127$ & $0.0176$ & $0.0092$ & $0.0215$ & $0.0115$ & $0.0324$\\ \cline{2-12}
& \multirow{2}*{\textbf{FTLSIN}} & \textbf{mean} & $\textbf{0.3530}$ & $\textbf{0.4374}$ & $\textbf{0.4980}$ & $\textbf{0.5519}$ & $\textbf{0.5876}$ & $\textbf{0.6179}$ & $\textbf{0.6580}$ & $\textbf{0.6712}$ & $\textbf{0.6967}$ \\
& & \textbf{variance} & $\textbf{0.0043}$ & $\textbf{0.0049}$ & $\textbf{0.0063}$ & $\textbf{0.0050}$ & $\textbf{0.0065}$ & $\textbf{0.0072}$ & $\textbf{0.0074}$ & $\textbf{0.0078}$ & $\textbf{0.0183}$\\ \hline
\multirow{8}*{Macro-F1} & \multirow{2}*{DeepWalk} & mean & $0.2523$ & $0.2667$ & $0.2768$ & $0.2945$ & $0.2935$ & $0.3077$ & $0.3101$ & $0.3294$ & $0.3359$ \\
& & variance & $0.0117$ & $0.0051$ & $0.0072$ & $0.0120$ & $0.0081$ & $0.0086$ & $0.0158$ & $0.0123$ & $0.0220$\\ \cline{2-12}
& \multirow{2}*{LINE} & mean & $0.3160$ & $0.2984$ & $0.3421$ & $0.3596$ & $0.4070$ & $0.4275$ & $0.4498$ & $0.4277$ & $0.4773$ \\
& & variance & $0.0113$ & $0.0127$ & $0.0144$ & $0.0249$ & $0.0382$ & $0.0548$ & $0.0383$ & $0.0302$ & $0.0486$\\ \cline{2-12}
& \multirow{2}*{Node2Vec} & mean & $0.4326$ & $0.4748$ & $0.5338$ & $0.5900$ & $0.6092$ & $0.6388$ & $0.6866$ & $0.6981$ & $0.6568$  \\
& & variance & $0.0147$ & $0.0156$ & $0.0153$ & $0.0153$ & $0.0290$ & $0.0314$ & $0.0202$ & $0.0572$ & $0.0261$\\ \cline{2-12}
& \multirow{2}*{\textbf{FTLSIN}} & \textbf{mean} & $\textbf{0.4662}$ & $\textbf{0.5094}$ & $\textbf{0.5747}$ & $\textbf{0.6354}$ & $\textbf{0.6557}$ & $\textbf{0.6863}$ & $\textbf{0.7377}$ & $\textbf{0.7488}$ & $\textbf{0.6908}$ \\
& & \textbf{variance} & $\textbf{0.0057}$ & $\textbf{0.0120}$ & $\textbf{0.0121}$ & $\textbf{0.0107}$ & $\textbf{0.0128}$ & $\textbf{0.0147}$ & $\textbf{0.0143}$ & $\textbf{0.0153}$ & $\textbf{0.0200}$\\ \hline
\end{tabular}
\end{table*}

DBLP dataset \cite{pan2016tri} (source network), consisting of bibliography data in computer science has been used widely in network and graph analysis \cite{jia:ICDM14,jia:TNNLS}. Each paper may cite or be cited by other papers, which naturally form a citation network. The network in this dataset abstracts a list of conferences from 4 research areas, \textit{i.e.,} database, data mining, artificial intelligence and computer vision.

CiteSeer-M10 dataset \cite{pan2016tri} (target network) is a subset of CiteSeerX data which consists of scientific publications from 10 distinct research areas, \textit{i.e.,} agriculture, archaeology, biology, computer science, financial economics, industrial engineering, material science, petroleum chemistry, physics and social science.

\subsection{Setups}
Our experiment evaluates the latent feature representations on standard supervised learning task: linear SVM classification. We choose the linear classifier instead of non-linear classifier or sophisticated relational classifiers in order to reduce the impact of complicated learning approaches on the classification performance. For evaluations, we randomly partition the dataset in the target domain into two non-overlapping sets for training and testing by 9 groups of training percents, $\{0.1,0.2,\cdots,0.9\}$. We repeat the above steps for 10 times and thus receive 10 copies of training data and testing data. The reported experimental results are the average of the ten runs and their variance.

\subsection{Benchmark Models}
\begin{figure*}[!t]
\centering
\subfigure[Random Walk on dblp]{\includegraphics[width=0.24\linewidth]{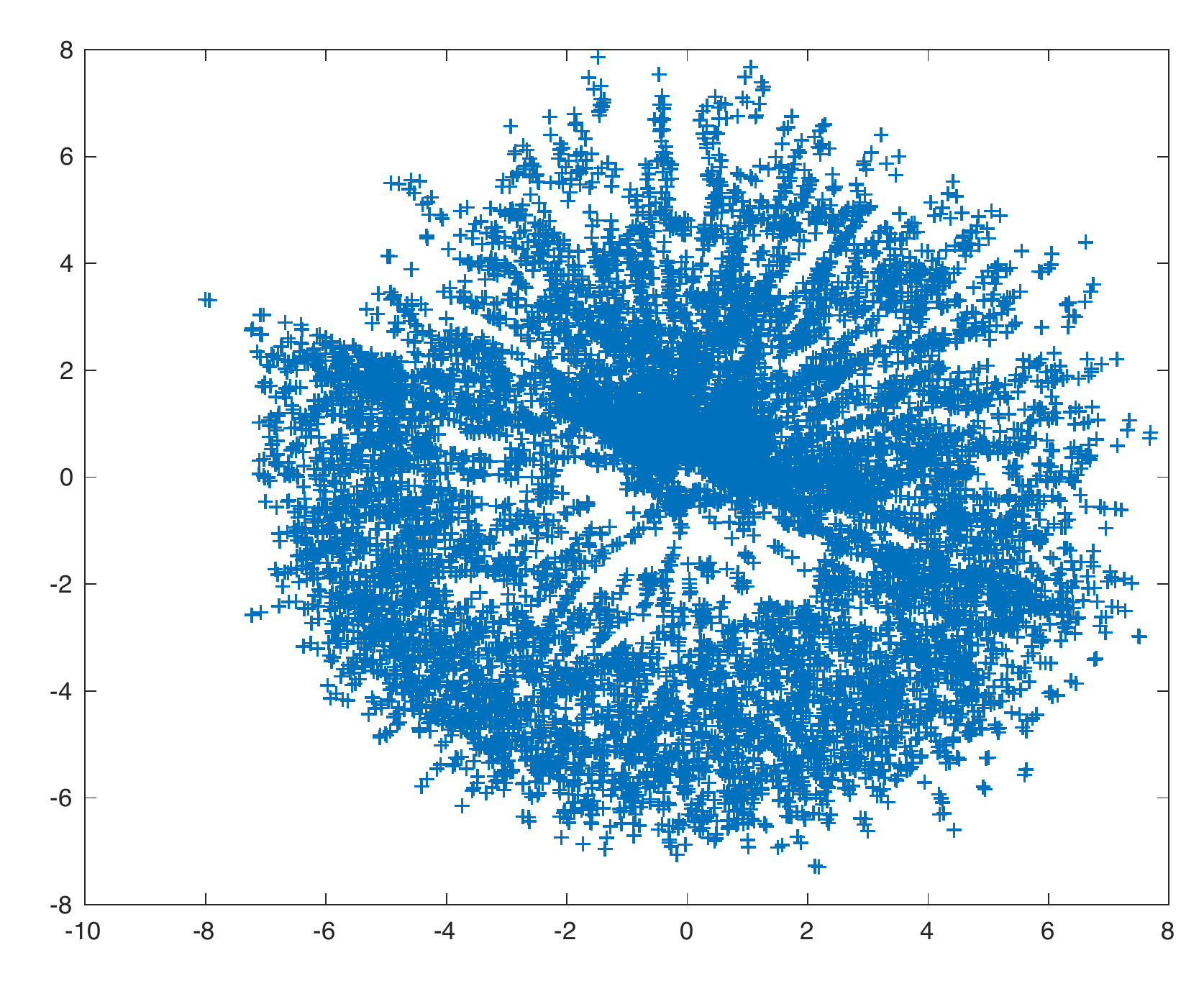}}
\subfigure[2-Layer Random Walks on M10]{\includegraphics[width=0.24\linewidth]{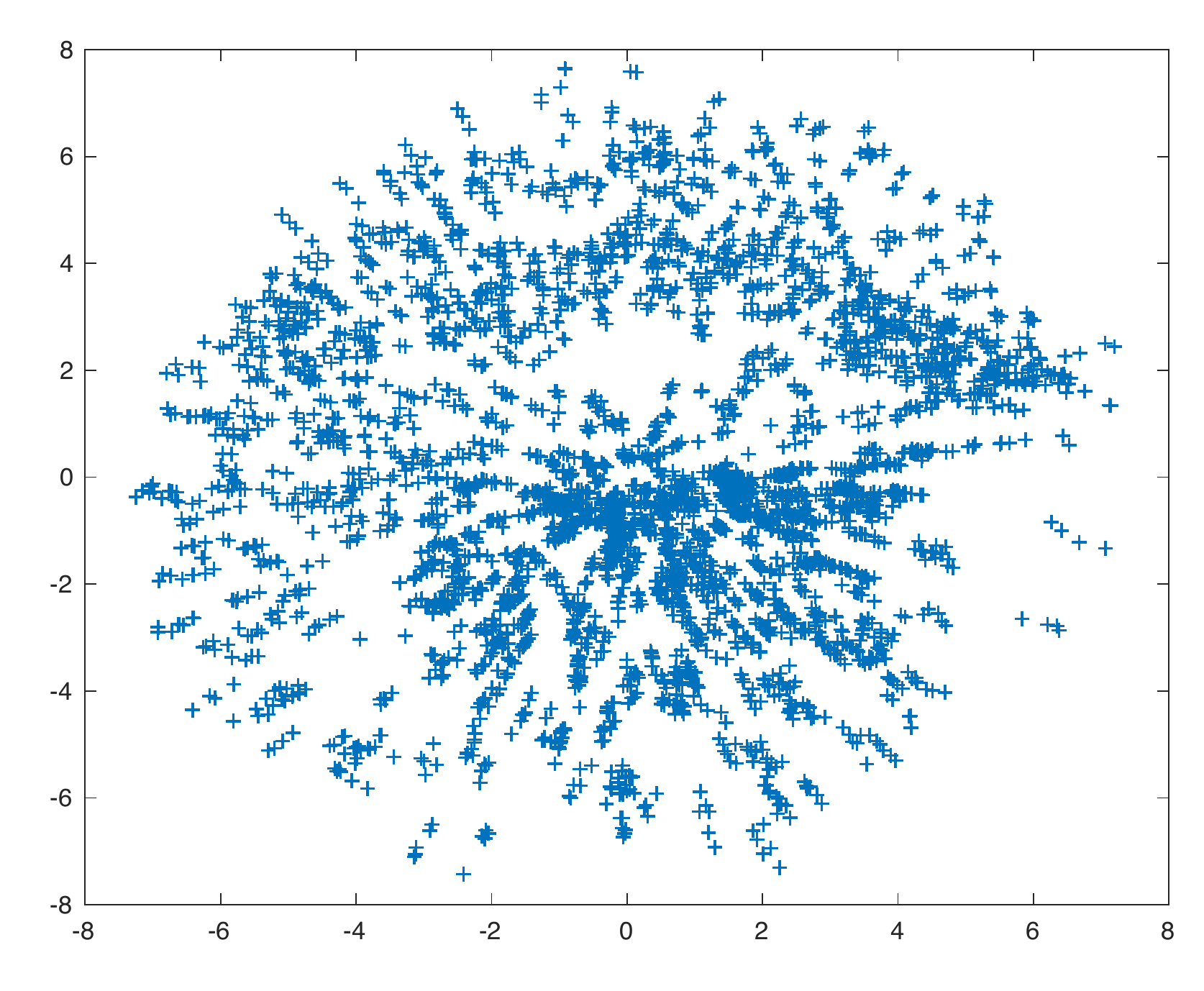}}
\subfigure[PCA on dblp]{\includegraphics[width=0.24\linewidth]{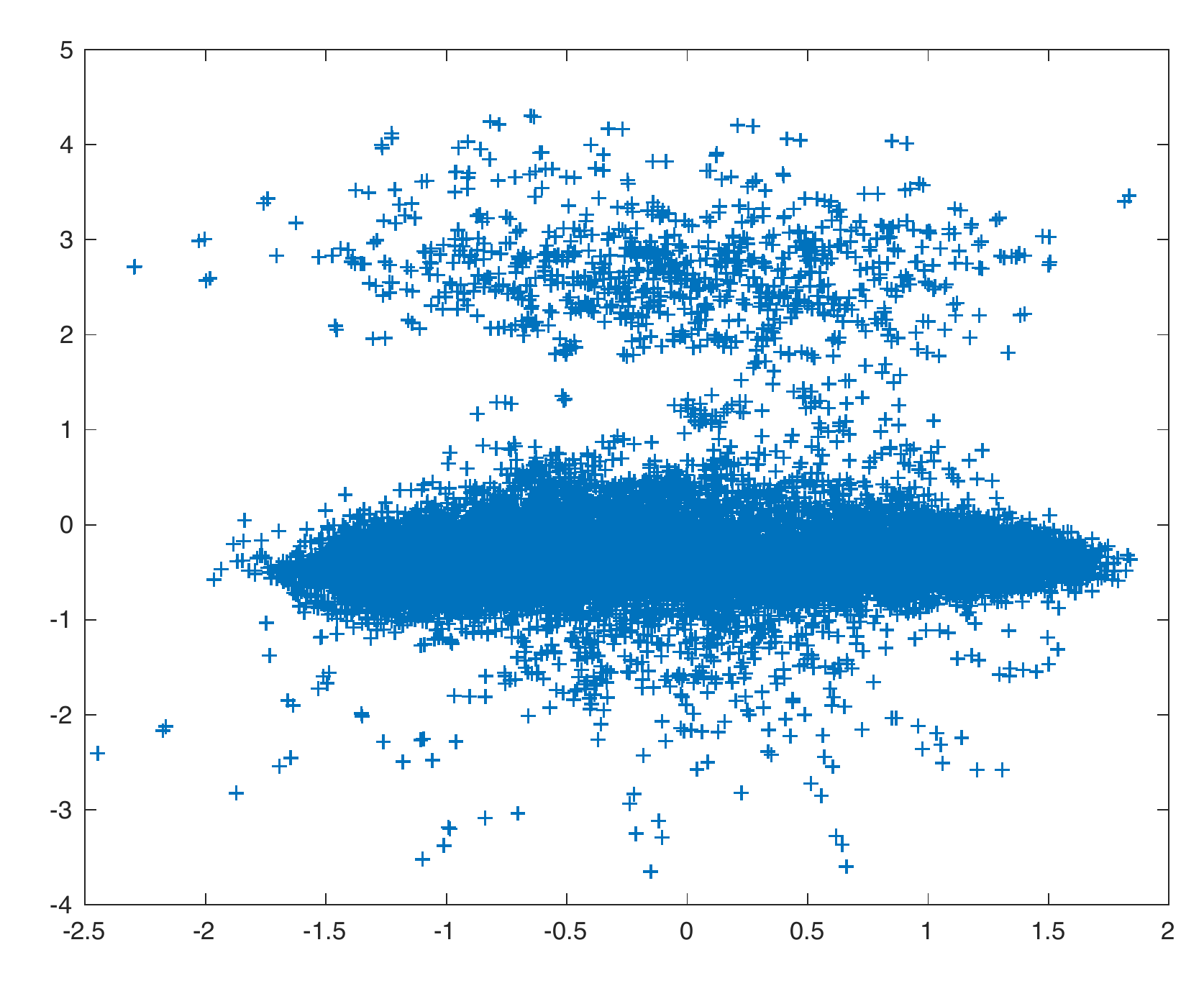}}
\subfigure[PCA on M10]{\includegraphics[width=0.24\linewidth]{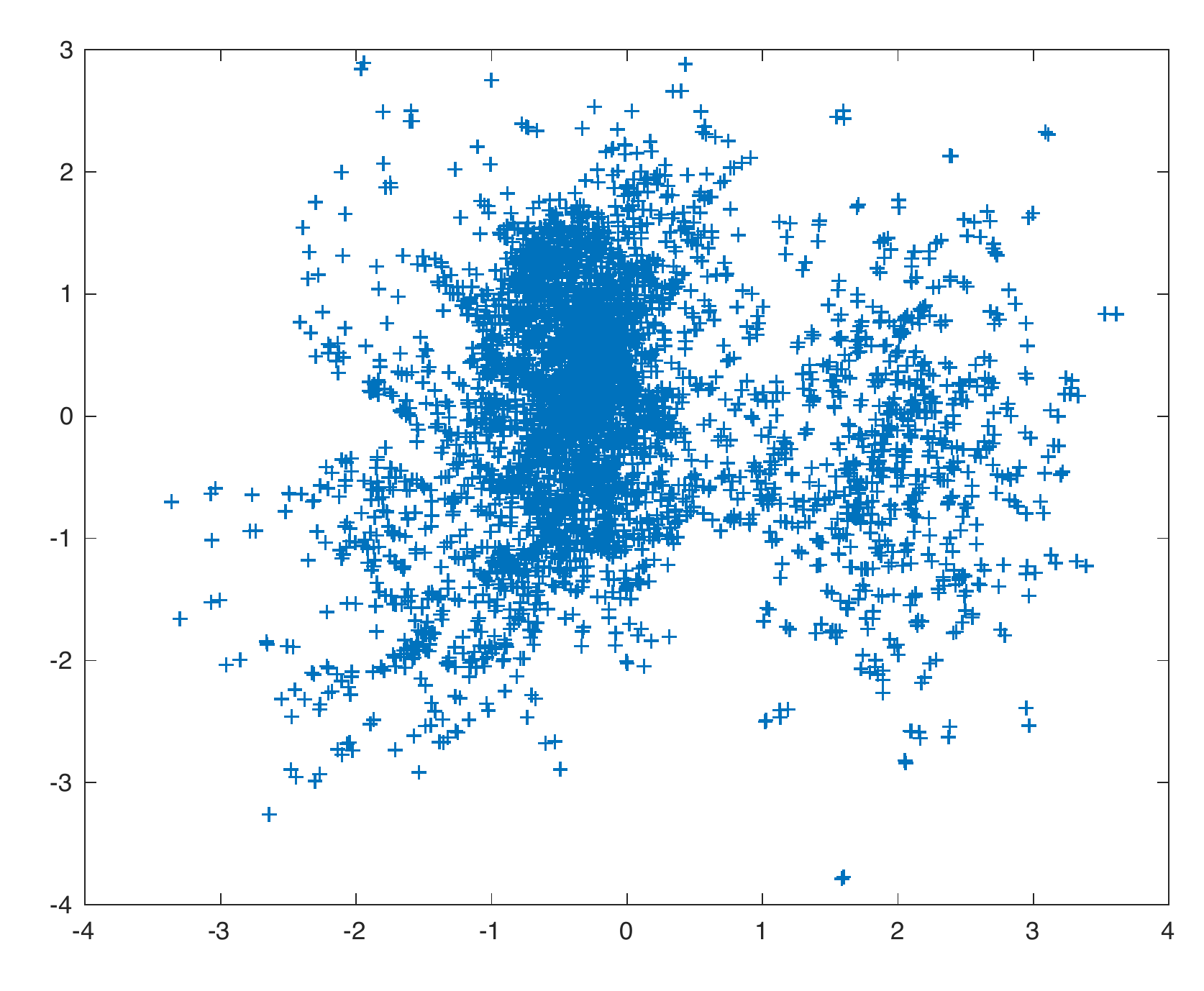}}
\subfigure[LLE on dblp]{\includegraphics[width=0.24\linewidth]{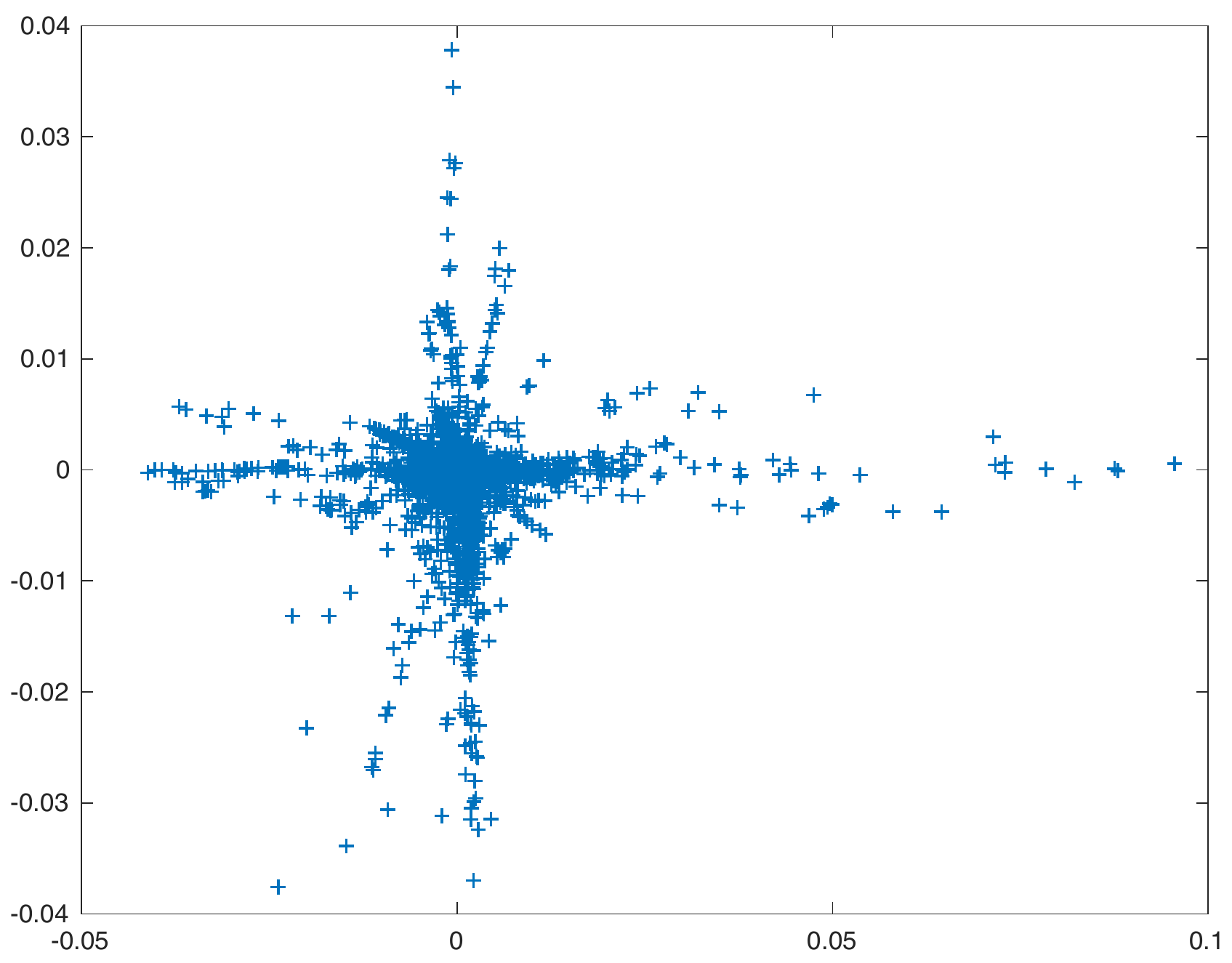}}
\subfigure[LLE on M10]{\includegraphics[width=0.24\linewidth]{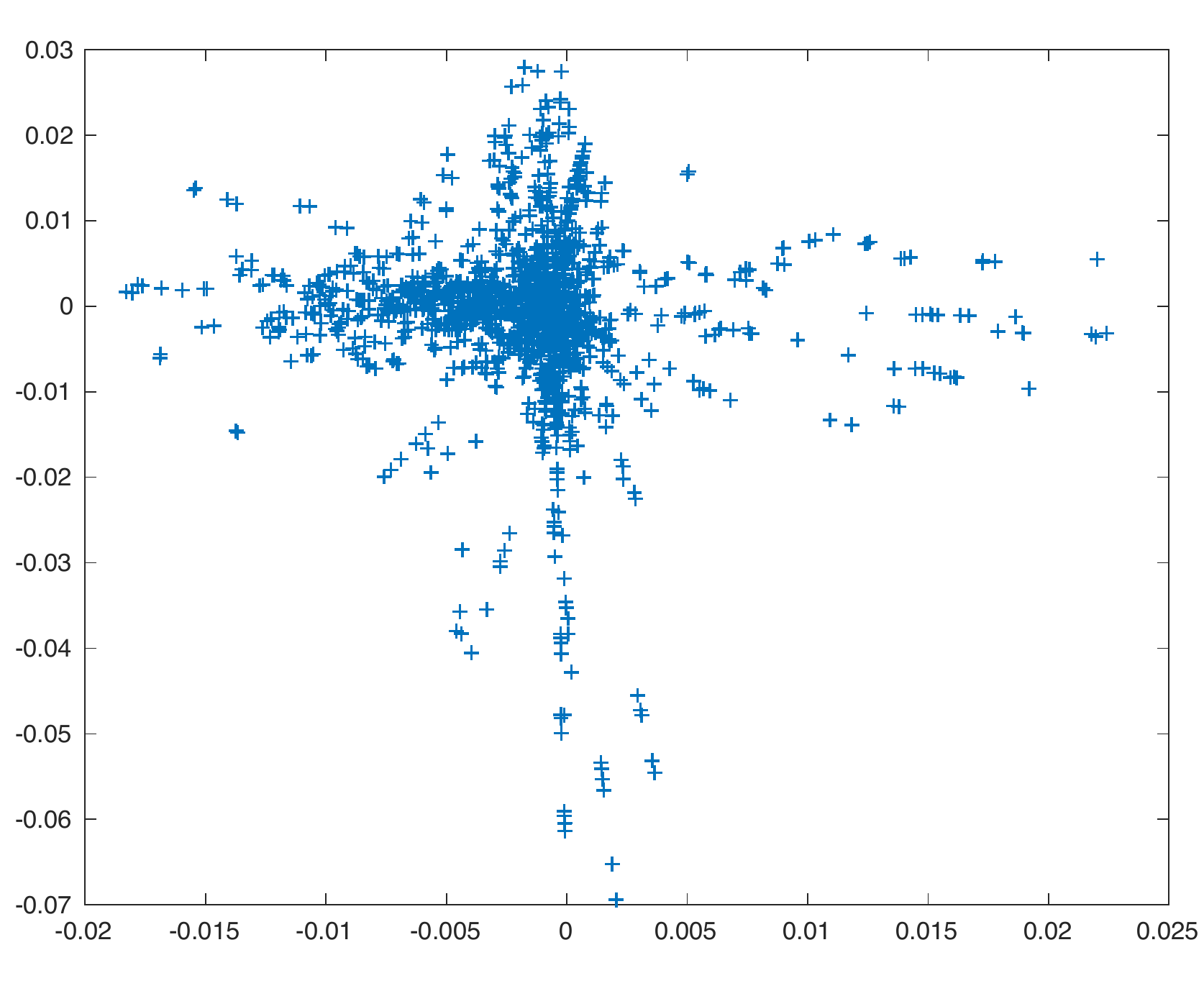}}
\subfigure[Laplacian on dblp]{\includegraphics[width=0.24\linewidth]{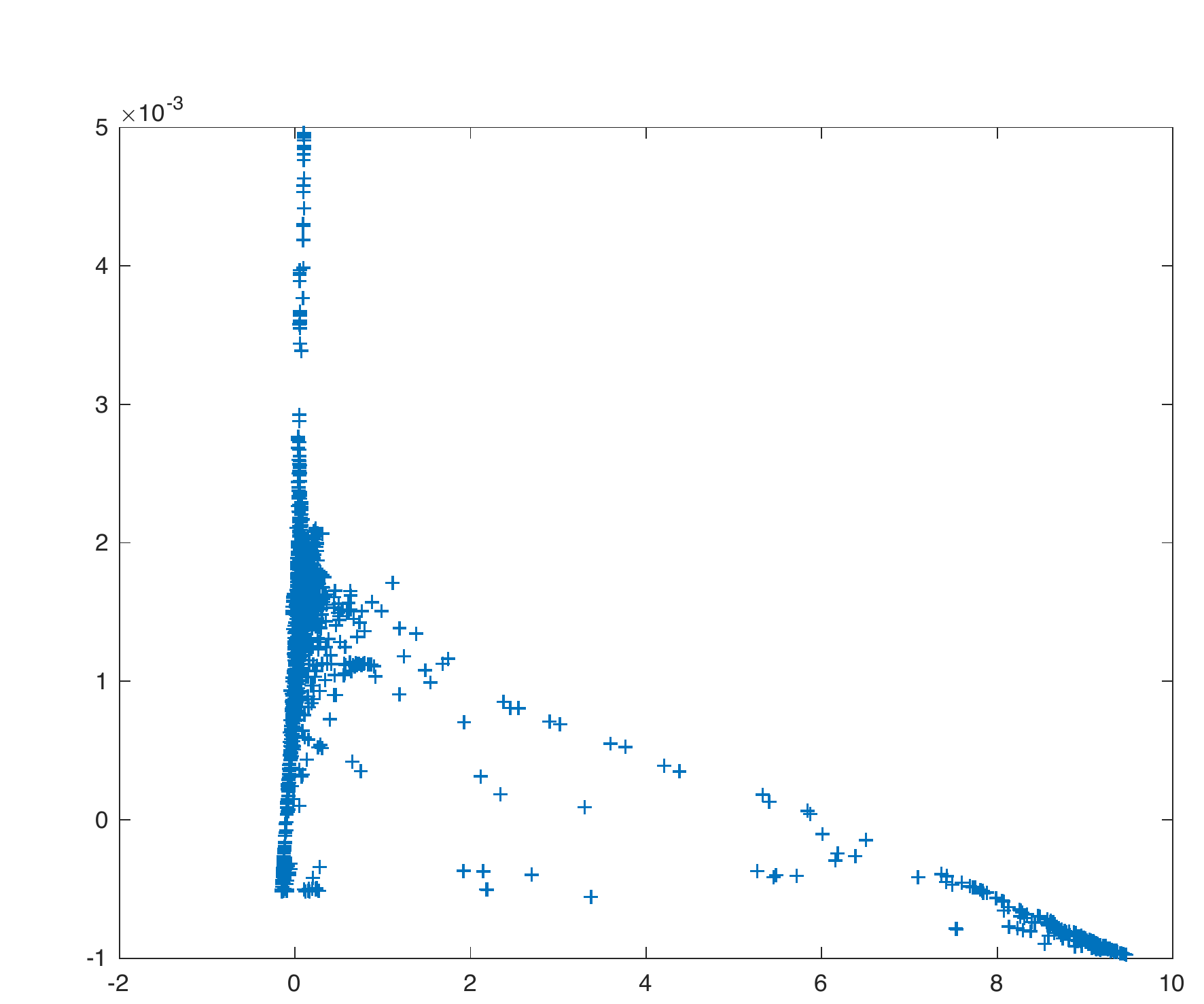}}
\subfigure[Laplacian on M10]{\includegraphics[width=0.24\linewidth]{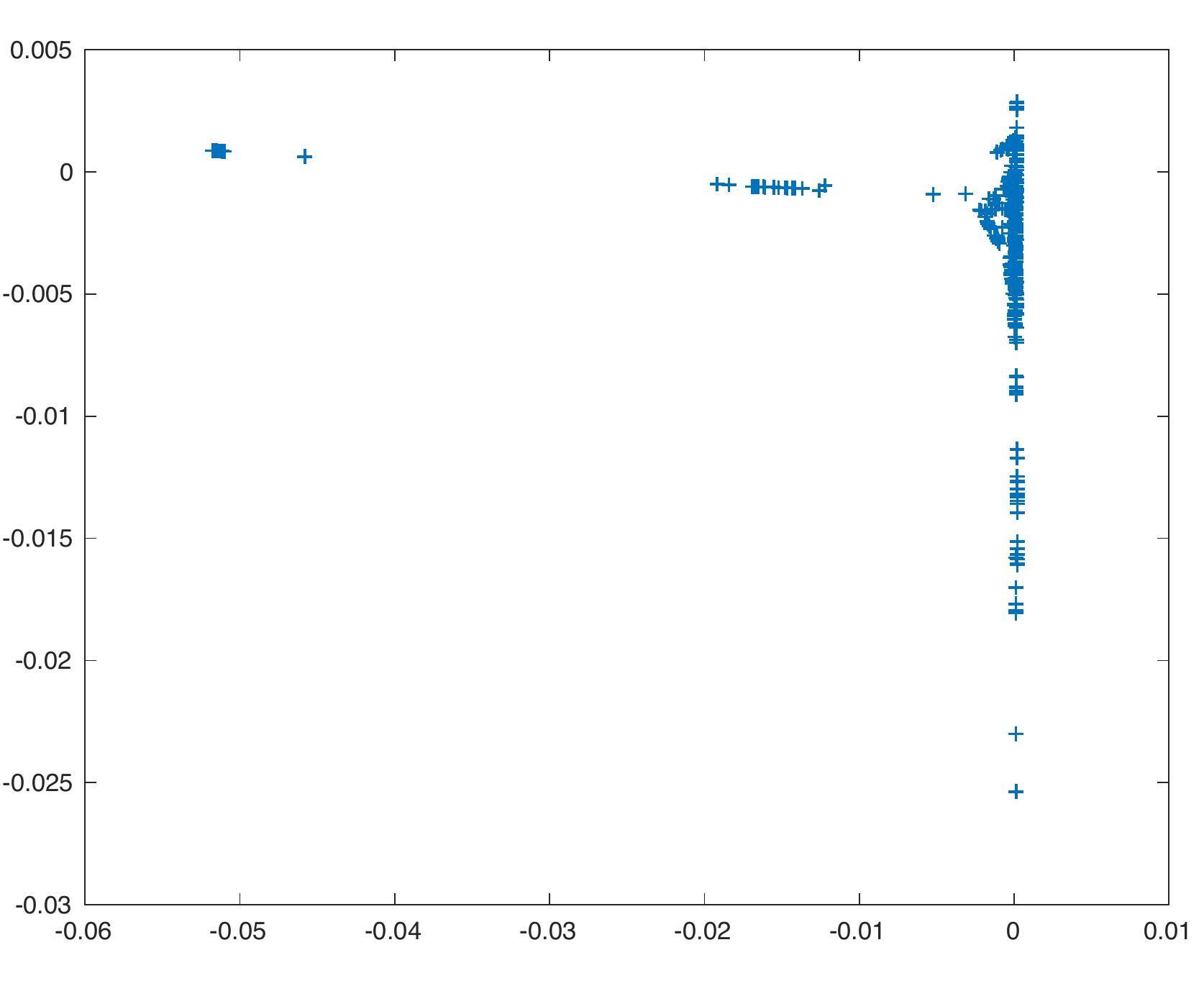}}
\caption{Network representations on the source network of dblp and on the target network of M10 in a 2-dimensional latent feature space.}
\label{fig:d2representation}
\end{figure*}

Fig. \ref{fig:powerlaw} and Fig. \ref{fig:powerlawwalks} show power law distributions \cite{adamic2000power} on the experiment datasets and their random walks, which obey the assumptions of the random walk that if the degree distribution of a connected graph follows a power law distribution, the frequency which the nodes appear in the short random walks will also follow a power law distribution \cite{perozzi2014deepwalk}.

We implement the following random walk based domain-specific network representation models for comparison. The benchmark models are applied with the FTLSIN Skip-gram for source networks in Eqs. (\ref{eq:source_skip-gram}), (\ref{eq:conditional_independence})-(\ref{eq:symmetry}).

1) \emph{DeepWalk} \cite{perozzi2014deepwalk}: This approach learns $d$-dimensional feature representations by simulating uniform random walks. The sampling strategy in DeepWalk can be seen as a special case of FTLSIN with bottom-layer random walk in $p=1$ and $q=1$.

2) \emph{LINE} \cite{tang2015line}: This approach learns $d$-dimensional feature representations in two separate phases. In the first phase, it learns $d/2$ dimensions by BFS-style simulations over immediate neighbors of nodes. In the second phase, it learns the next $d/2$ dimensions by sampling nodes strictly at a 2-hop distance from the source nodes.

3) \emph{Node2Vec} \cite{grover2016node2vec}. This approach learns $d$-dimensional feature representations by BFS-style simulations over immediate neighbors of nodes. The sampling strategy in Node2Vec is also a special case of FTLSIN with bottom-layer random walk in $p=1$ and $q=1$.

\subsection{Parameters Setting}

The parameter settings used for FTLSIN are in line with typical values used for DeepWalk, LINE and Node2Vec. Specially for source and target networks, we set the dimensions of feature representation as $d=128$, set the walk length as $l=80$, set the number of walks of every source node as $k=10$, and set the window size as $r=10$. In this way, the total number of walks over a input network is $w=SampleSize\times k$, and the shape of walk sets are in $w\times l$. The parameters of search bias $\alpha$ is set in $p=1$, $q=1$.

\subsection{Experimental Results}

The node feature learning by network representations are input to a one-against-all linear SVM classifier \cite{hsu2002comparison}. We use Macro-F1 and Micro-F1 for comparing performance and the results are shown in Table \ref{tab:result_target}. These two measures are popular just like the classification accuracy performance in data mining areas \cite{Wu:2012:HDK,6707028,Wu:2014:NBP:2628618.2628680}.

\textbf{Representation Analysis.} Fig. \ref{fig:d2representation} (a) illustrate the feature spaces of dblp by FTLSIN bottom layer random walk, Fig. \ref{fig:d2representation} (b) illustrate the feature spaces of dblp by FTLSIN 2-layer random walks. These two illustrations show almost the same distributions in feature spaces and get good mappings in a low dimension than PCA, LLE and Laplacian based network representaitons.

\textbf{Effectiveness of search priority in random walks.} In Table \ref{tab:result_target}, DeepWalk and LINE show the worse performance than our FTLSIN and Node2Vec, which can be explained by its inability to reuse samples, a feat that can be easily done using the random walk. The outstanding of Node2Vec among benchmark models indicates the exploration strategy is much better than the uniform random walks learned by DeepWalk and LINE. Meanwhile, the poor performance of DeepWalk and LINE is mainly because the network structure is rather sparse, with noises and only contains limited information. FTLSIN and Node2Vec are both good performing on M10 network with above advantages, as parameter of search bias $\alpha$ adds the flexibility in exploring local neighborhoods prior to global network.

\textbf{Importance of information from source domain.} Table \ref{tab:result_target} shows that FTLSIN outperforms the domain-specific benchmark models, which uses topological information from the source domain to learn the network representation in the target domain. When we add a top-layer in 2-layer random walks, the information in the source network are transferred to the source network by adjusting the weights on the edges of the target network.

\section{Conclusion and Future Work}\label{sec5}

In this paper, we propose a solution for a new scenario in network representation, that is transferring structures across networks with 2-layer random walks. Our framework effectively improves the performance of latent feature learning in large-scale citation networks. Meanwhile, it reduces learning difficulties of data sparsity and noises. Future works include FTLSIN with multiple labels and deep network representation \cite{7453156,7890384}.

\def\IJCNN{\it International Joint Conference on Neural Networks\rm }

\bibliographystyle{IEEEtran}
\bibliography{reference}

\end{document}